\documentclass[twocolumn,showpacs,pre]{revtex4}

\begin{document}

\title{General relation between drift velocity and dispersion\\
       of a molecular motor}

\author{Zbigniew Koza}
\email{zkoza@ift.uni.wroc.pl} \affiliation{
             Institute of Theoretical Physics, University of
             Wroc{\l}aw, pl.\ Maxa Borna 9, 50204 Wroc{\l}aw,
             Poland
           }


\begin{abstract}
We model a processive linear molecular motor as a particle diffusing in a
one-dimensional periodic lattice with arbitrary transition rates between its
sites. We present a relatively simple proof of a theorem which states that
the ratio of the drift velocity $V$ to the diffusion coefficient $D$ has the
upper bound $2N/d$, where $N$ is the number of nodes in an elementary cell
and $d$ denotes its length. This relation can be used to estimate the minimal
value of internal states of the motor and the maximal value of the so called
Einstein force, which approximately equals to the maximal force exerted by a
molecular motor.

\end{abstract}

\pacs{87.16.Nn. 05.10.Gg, 05.40.-a}

\maketitle


\section{Introduction}
\label{Introduction} Molecular motors are enzymes mediating
intracellular transport and cellular motion in living cells. The most
important examples are kinesins (responsible for transport of organelles),
myosins (which drive muscle contraction) and dyneins (involved in cellular
locomotion)\cite{HowardBook}.

Based on their function and structure, molecular motors can be divided into
several families. Here we will focus on {\em processive linear} motors. A
motor is linear if it moves along a complementary protein fiber (other types
of motors include protein pumps, involved in transport across membranes, and
rotary motors).  A motor is processive if it can travel a long distance
before it detaches from the filament.

A typical example of a processive linear motor is kinesin. It is a
mechanoenzyme which transduces the chemical energy of hydrolysis of ATP
molecules into mechanical work, which is then used to move vesicles and
organelles along microtubules. Each step is powered by hydrolysis of one ATP
molecule. The main theoretical challenge is to explain how the chemical
energy of ATP is transformed into practically unidirectional motion of
protein motors.

Theoretical modeling of the motion of motor proteins has been mainly based
on two approaches. The first one, called a thermal ratchet model, regards a
motor protein as a Brownian particle diffusing in several periodic,
asymmetric potentials which are alternatively switched on and off in
stochastic time intervals \cite{Julicher-RMP97}. In the following we employ
the second approach, which is based on an assumption that the motion of a
motor can be modeled as a sequence of transitions between discrete
mechano-chemical states of the motor, which are represented as nodes of a
linear lattice \cite{FisherPNAS,KoloJChP2000,FischerPNAS01}. A segment of a
microtubule corresponds to an elementary cell of the lattice, and the number
of sites in an elementary cell, $N$,  is equal to the number of different
internal states in a full mechano-chemical cycle of a motor. For kinesin the
lattice constant $d=8$ nm \cite{HowardBook,Coppin,Visscher99} and $N \ge 4$
\cite{FischerPNAS01}.

Any theoretical model must account for experimental results. Here we will be
interested in an approach which attempts to determine internal properties of
molecular motors based on the statistical analysis of their motion as
observed in experiments. The fundamental quantities employed in this method
are the mean velocity $V = \left<x(t)\right>/t$ and the dispersion $D=
\langle(x (t) - \langle x(t)\rangle)^2 \rangle /2t$. Having determined these
two parameters (in the limit of large $t$), one can find the so called {\em
randomness} $r$ defined simply as \cite{SvobodaPNAS94, FischerPNAS01}
\begin{equation}
 \label{defR}
  r \equiv 2D/V d.
\end{equation}
This quantity can be used to estimate the number of internal states of a
molecular motor. The idea is simple: if transitions constitute a Poisson
unidirectional process with exactly the same transition rates $k$, then $r =
1/N$ \cite{SvobodaPNAS94}. This idea was extended by Fisher and Kolomeisky
\cite{FisherPNAS,FischerPNAS01}, who made a conjecture that in a general case
of a linear chain with bidirectional  Poissonian transitions there is
\begin{equation}
 \label{NGE1R}
    N \ge 1/r,
\end{equation}
which enables one to determine the minimal number of internal states in a
full mechano-chemical cycle of a motor.

Another important quantity which can be directly related to $V$ and $D$ is
the so called Einstein force
\begin{equation}
    F_E \equiv k_{B}T \frac{V}{D},
\end{equation}
where $T$ is temperature and  $k_{B}$ denotes the Boltzmann constant. This
quantity is a linear-response estimation of the force exerted by a molecular
motor \cite{FisherPNAS,FischerPNAS01}. Conjecture (\ref{NGE1R}) can be now
written as
\begin{equation}
  \label{hypo} F_{E} \le 2 k_{B}T \frac{N}{d}.
\end{equation}
The maximal value of the Einstein force is thus a simple function of three
parameters: temperature $T$, a step length $d$ and the number $N$ of
different internal states of a motor.

In a recent paper \cite{KozaPRE01} we have given a rigorous proof of relation
(\ref{NGE1R}) for linear chains with arbitrary Poisson transitions between
nearest-neighbor sites. However, that proof is rather lengthy and
complicated. Here we present a different argument justifying (\ref{NGE1R}) and
(\ref{hypo}). Not only is it much simpler, but it is also more general, as it
can be used for systems with transitions between arbitrary lattice sites
(provided the lattice is periodic).

The paper is organized as follows. In Section \ref{Model} we present a
mathematical definition of the model. Section \ref{Proof} contains the proof
of equation (\ref{NGE1R}), and Section \ref{Discussion} is devoted to the
discussion of our results.


\section{Model}
\label{Model}
We consider a one-dimensional lattice with its sites located at $x_n$,
$n\in{\cal Z}$.  At time $t=0$ we put a particle at site $x_0=0$ and let it
jump between the nearest-neighbor lattice sites. Transitions are assumed to
represent a Poisson process governed by the master equation
\begin{equation}
\label{CMEL}
   \frac{\partial P(n,t)}{\partial t}
   =    k^{+}_{n-1} P(n-1,t) + k^{-}_{n+1} P(n+1,t)
                   - (k^{+}_{n} + k^{-}_{n}) P(n,t),
\end{equation}
where $P(n,t)$ denotes the probability of finding the particle at site $x_n$
at time $t$ and $k^\pm_n \ge 0$ are the (constant in time) transition rates
from a site $x_n$ to $x_{n\pm 1}$. We assume that the system is periodic in
space with a period  $N\ge 1$ and a lattice constant $d > 0$. We do not
demand that the distances between consecutive lattice sites, $x_{n+1} -
x_n$, should be all equal to each other. Our goal is to prove (\ref{NGE1R})
for any choice of $N$, $k^{+}_n$ and $k^{-}_n$.


\section{Proof of relation (\protect\ref{NGE1R})}
\label{Proof}

Relation  (\ref{NGE1R}) is equivalent to
\begin{equation}
   \label{HYPO2}
    D \ge \frac{Vd}{2N} .
\end{equation}
We will prove this inequality by fixing the values of $V$, $d$ and $N$ and
finding the minimal value of $D$ as a function of $d$ and transition rates
$k^{\pm}_n$, $n=1,\ldots,N$. Since $V/d$ and $D/d^2$ depend only on
transition rates $k^{\pm}_n$  \cite{Koza}, the problem of proving
(\ref{HYPO2}) reduces to the one of minimizing a rather complicated function
of $2N$ nonnegative variables $k^{\pm}_n$.

Probably the simplest proof of inequality (\ref{HYPO2}) is based on a very
general and nontrivial property of diffusive lattice systems \cite{Koza}.
Suppose we have a diffusion process on a periodic lattice with Poissonian
transitions between its sites and let $k_{mn}\ge 0$ denote the transition
rate from site $n$ to $m$. There exists a time interval $\tau$ such
$k_{mn}\tau \le 1$ for all $n$ and $m$. We can thus define another
stochastic process on this lattice, in which transitions can occur only at
regular intervals $\tau$ and are characterized by {\em probabilities}
$k_{mn} \tau$ ($k_{mn} \tau$ is the probability of jumping from $n$ to $m$).
It has been rigorously shown \cite{Koza} that the drift velocity $V$ and
dispersion $D$ calculated in the original, continuous-time system are
related to the drift velocity $V^D$ and dispersion $D^D$ of its
discrete-time counterpart through general formulae
\begin{equation}
 \label{VD}
  V = V^{D}, \quad D = D^{D} + \frac{\tau}{2}V^2.
\end{equation}
The problem of finding the combination of transition rates $k^\pm_n$ which
minimize the value of dispersion $D$ for some fixed values of $V$ and $d$ in
the continuous-time process is therefore equivalent to the same problem posed
for the corresponding discrete-time system. However, the latter problem is
{\em much} simpler. By definition  dispersion $D^D$ cannot be negative.
Moreover, $D^D=0$ for two particular choices of transition probabilities
$k^{\pm}_n\tau$: either when $k^{-}_n \tau = 0$ and  $k^{+}_n \tau = 1$ or
when $k^{-}_n \tau = 1$ and $k^{+}_n \tau = 0$. In these two cases the
motion of a diffusing particle is actually unidirectional and {\em
deterministic} --- hence the value of dispersion $D^{D} = 0$. Clearly,
$D^{D}$ assumes the minimal possible value {\em only} in these two particular
''degenerated'' cases where the stochastic, bidirectional process turns into
a unidirectional, deterministic one. Non-vanishing transition rates in the
corresponding continuous-time process are equal $1/\tau$, and in this case
 $V = d/N\tau$ and $D=(d/N)^2/2 \tau = Vd/2N$
\cite{Koza}. The minimal value of dispersion $D$ is thus equal to $Vd/2N$,
which completes the proof of (\ref{HYPO2}).

It is worth noting that not only have we just proved inequality
(\ref{HYPO2}), but but we also showed that $V/D$ assumes its maximal value,
$2N/d$, if and only if all transitions are unidirectional and of the same
magnitude, i.\ e.\ when for any $1\le n,m \le N$ there is
$$
  k^{+}_n = k^{+}_m > 0,\; k^{-}_n = 0
  \quad \mbox{or} \quad
  k^{-}_n = k^{-}_m > 0,\; k^{+}_n = 0.
$$

It is also interesting to note that in our present approach we almost did
not employ the assumption that only transitions between nearest-neighbor
lattice sites are allowed. The same line of reasoning could be applied for
more general systems. The only difference would be that in a general case
there would be several choices of transition rates for which the auxiliary
discrete-time process is deterministic. Some of them could be characterized
by different values of $N$ and we would have to plug  the smallest of them
in Eq.\ (\ref{HYPO2}). However, in the case of molecular motors such
complicated models do not seem to be appropriate, as we expect that their
behavior is governed by the most probable reaction path \cite{HowardBook}.


\section{Conclusions}
\label{Discussion}

We have rigorously proved that in diffusive processes on periodic lattices
with Poissonian transitions between its nodes the value of diffusion
coefficient $D$ is bounded from above by $Vd/2N$, where $V$ is the mean
drift velocity, $d$ is the length of the elementary cell, and $N$ denotes
the number of sites in the elementary cells. The present approach is much
simpler than our previous method employed in Ref.\ \cite{KozaPRE01}. It is
also more general, as it can be applied to  arbitrary lattice systems, while
our previous approach was applicable only to linear lattices with transitions
restricted to the nearest-neighbor sites.

Our result has interesting applications in the theory of molecular motors.
First of all it implies that by measuring experimentally the randomness $r
\equiv 2D/V d$ one can determine the minimal number $N$ of internal states
of a motor in its full mechano-chemical cycle. As a matter of fact, this
approach was already employed in Ref.\ \cite{FischerPNAS01}, where it was
found that for kinesin $N \ge 4$. Our result gives firmer basis for this
kind of argumentation.

Our result implies also that the maximal value of the Einstein force,
$F_{E}$, is $2k_{B}T N/d$, where $T$ is the temperature and $k_{B}$ -- the
Boltzmann constant. We showed that this maximal value is attained if and
only if all transitions are unidirectional and of the same magnitude.

Although our approach is valid for any value of $N$ and an arbitrary choice
of transition rates, in the context of molecular motors it still refers to a
simplified situation. First, we assumed that the motion of a motor protein
can be reduced to diffusion of a Brownian particle on a linear chain of
lattice nodes, and other topologies deserve at least equal attention
\cite{Julicher-RMP97,KoloJChP2000,MotorsB,Lipovsky}. Second, we assumed that
transitions constitute a Poisson process. Although this hypothesis is
confirmed by recent experiments on myosin \cite{Rief}, it is well known that
transition rates with other probability density functions may lead to
completely different upper bounds for dispersion $D$ (and hence the upper
bound for $F_{E}$ and the lower bound for $r$ may also change)
\cite{FisherPNAS,KoloJChP2000,FischerPNAS01}. Nevertheless, the force
calculated using all these approximations ($4.3$ pN) agrees quite well with
the experimental value of the stalling force $F_{\mathrm s}$ for kinesin
(different experimental techniques yielded $F_{\mathrm s} \sim 4-8$ pN
\cite{HowardBook}). Further work is, of course, required to clarify the
relevance of the above-mentioned problems.


\section*{Acknowledgments}

Support from the Polish KBN Grant Nr 2 P03B 119 19 is gratefully
acknowledged.



\end{document}